# Direct Gaussian Process Quantile Regression using Expectation Propagation

**Alexis Boukouvalas**  boukouva@aston.ac.uk
**Remi Barillec**  r.barillec@aston.ac.uk
**Dan Cornford**  d.cornford@aston.ac.uk
Aston University, Birmingham B4 7ET, UK

## Abstract

Direct quantile regression involves estimating a given quantile of a response variable as a function of input variables. We present a new framework for direct quantile regression where a Gaussian process model is learned, minimising the expected tilted loss function. The integration required in learning is not analytically tractable so to speed up the learning we employ the Expectation Propagation algorithm. We describe how this work relates to other quantile regression methods and apply the method on both synthetic and real data sets. The method is shown to be competitive with state of the art methods whilst allowing for the leverage of the full Gaussian process probabilistic framework.

## 1. Introduction

Quantile regression has been applied in a variety of domains and for different purposes (Yu et al., 2003). Applications include medical reference charts, survival analysis, economics and the detection of heteroscedasticity. Quantile regression allows a comprehensive analysis of the relationships between a response, $y$, and input variables, $x$. In traditional regression analysis there is often an implicit assumption that any uncertainty in the learned model is a result of incomplete knowledge of an underlying deterministic function due to incomplete, noisy observations. However, quantile regression is most relevant when the response is likely to be subject to variability or intrinsic randomness, such as might occur in population or meta-studies, regression modelling where not all relevant inputs are available or considered, or when modelling the output of a stochastic simulation.

Two main approaches to quantile regression have been described in the literature: the Estimating Equation (EE) approach and inverting a Cumulative Distribution Function (CDF). The EE approach is based on directly modelling the quantile function, learning the parameters by minimising an appropriate loss function. The CDF approach is based on estimation of the CDF of the response and inverting this to obtain the desired quantiles. This differentiation is akin to the difference between discriminatory (EE) and generative (CDF) models in classification.

In the quantile regression setting, the generative case corresponds to estimating the full conditional CDF $F(y|x)$ and then inverting it to obtain specific quantiles. This *model-based* approach allows for a natural Bayesian formulation. Taddy & Kottas (2010) provide a clear overview of options for quantile regression, and propose a model based on a non-parametric Dirichlet process prior to construct a flexible joint model for the response and inputs, conditioning this on inputs to obtain the required conditional response distribution. Chen & Muller (2012) develop a non-parametric CDF approach based on computing indicator functions which are smoothed using a kernel spectral decomposition to form the full conditional density model, and thus determine the quantile functions. A related approach in the field of Geostatistics is known as indicator cokriging (Pardoiguzquiza & Dowd, 2005) where a Gaussian process is used to estimate a discretised approximation to the CDF. All of these methods require carefully designed Markov Chain Monte Carlo (MCMC) inference methods, and require significant computational effort, making their application to problems with many inputs infeasible. Alternatively for a Gaussian posterior model the inversion of the CDF can be done analytically to retrieve quantile functions as demonstrated in Quadrianto et al. (2009).

The major advantage of the CDF approach is that an appropriate likelihood function is defined and the joint estimation of the quantiles means order violations (quantile crossings) are not possible by construction. However in scenarios where the interest is in the specification of one, or a small set of quantiles, intuitively it seems unnecessary to attempt to describe the entire conditional distribution. The direct EE approach may also be more appropriate in application domains such as real-time systems where inference time is critical as it allows for faster computation using simpler models than the CDF approach.

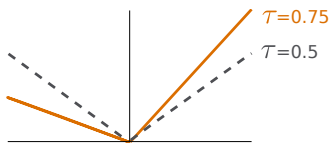

Figure 1. Tilted loss function for two different quantiles, $\tau$.

The EE approach (Koenker, 2005) can be seen as akin to directly constructing a decision boundary to separate the classes. In the quantile regression case a loss function is minimised to obtain the quantiles of interest directly. Various EE approaches to quantile regression exist. The frequentist interpretation minimises an empirical risk function, related to the tilted loss function (Figure 1) given by

$$L_\tau(y - y^*) = \begin{cases} \tau(y - y^*) & \text{if } y \geq y^* \\ -(1-\tau)(y - y^*) & \text{if } y < y^* \end{cases}$$

which has been shown to consistently estimate the $\tau$'th quantile. The tilted loss is also known as the pinball loss (Takeuchi et al., 2006).

Many approaches use linear in parameter, or spline models (Koenker, 2005). These approaches are consistent and produce classical estimates for the quantile functions, which can also incorporate the use of simplex methods or post processing to ensure no order violations for multiple quantiles (Koenker, 2005). Many papers have also attempted to provide a 'Bayesian' version of the EE approach to quantile regression. These exploit the association between the tilted loss function and the Asymmetric Laplace Distribution (ALD), as explained in more detail in Section 2.

Yu & Moyeed (2001) develop a 'Bayesian' linear model for quantile regression assuming an ALD likelihood, but care must be taken in the interpretation of the posterior distribution in the typical case that one does not believe the errors on the response actually follow an ALD. The use of the ALD remains common (Yue & Rue, 2011; Lum & Gelfand, 2012), and while the mode of the solution can be shown to be consistent with the true quantile, the uncertainty on the quantiles has no clear interpretation. In Lum & Gelfand (2012) a conditional Gaussian representation of the ALD is used to incorporate spatially dependent errors. Inference in this model is accomplished via MCMC. Their approach is quite similar to what is proposed in this paper in that spatial dependence is modelled via a stochastic process. Yue & Rue (2011) propose a similar model where a Gauss-Markov random field model is used to address spatial (input) dependency and both iterated nested Laplace approximations and MCMC are used for inference.

Our focus in this paper is on quantile regression where the conditional quantile functions are of interest. Our contribution consists of presenting a novel method for quantile regression which uses approximate inference methods to improve efficiency. We place a Gaussian Process (GP) prior on the quantile regression function similarly to Lum & Gelfand (2012) and directly minimise the expected tilted loss using an Expectation Propagation (EP) approach (Minka, 2001). Further we clarify the justification of the EE approach which has been used by a variety of authors and show that although not truly a Bayesian approach, a decision theoretic grounding is possible.

The paper is structured as follows. In Section 2 we present our approach. A simulation study on both synthetic and real data is presented in Section 3. A discussion of the results and future extensions is given in Section 4.

## 2. GP Quantile Regression using EP

Our model, which we term QGP-EP, places a GP prior on the space of quantile regression functions[1]. The training of the model proceeds by directly minimising the expected loss or maximising an equivalent utility function. The latter is found to correspond to the ALD which has been widely used in direct quantile estimation. The expected utility turns out to not be analytically tractable and we employ the EP algorithm to perform the integration. A high level description of the algorithm is given in Algorithm 1. We conclude by discussing how hyper-parameter estimation and prediction are accomplished.

Minimising the expected tilted loss

$$\underset{y}{\operatorname{argmin}} \int L(y_*, y) p(y_* | x_*, \mathcal{D}) \, \mathrm{d}y_* \quad (1)$$

leads to the $\tau$'th quantile of $p(y_*|x_*, \mathcal{D})$. This is re-ferred to as the expected quantile risk in Takeuchi et al. (2006). For more details on the optimality conditions for different loss functions see Berger (1985). If we take the exponent of the negative of the tilted loss and normalise, we have the ALD (Yu & Zhang, 2005). The density function is:

$$\mathcal{L}(t|\mu, \sigma, \tau) = \frac{\tau(1-\tau)}{\sigma} \exp\left[-\frac{t-\mu}{\sigma}(\tau - I(t \leq \mu))\right]. \quad (2)$$

---

[1] Code is available at http://wiki.aston.ac.uk/AlexisBoukouvalas.

The parameter $\tau \in [0,1]$ controls the skewness of the distribution. For $\tau = 0.5$ we retrieve the Laplace distribution. The mean $\mu$ can take any real value and the standard deviation has to be positive $\sigma > 0$. The indicator function $I(t \leq \mu)$ is 1 if the condition is true, 0 otherwise. We can therefore define a utility:

$$\mathcal{U}_\tau(\mathbf{y}, \mathbf{q}) = Z \exp[-L_\tau(\mathbf{y}, \mathbf{q})]. \quad (3)$$

where $\mathbf{q}$ is the predicted value of the $\tau$ quantile, $\mathbf{y}$ the observations and $Z$ the normalisation. If we take the common assumption that the utilities are independent for each observation, we have:

$$\mathcal{U}_\tau(\mathbf{y}, \mathbf{q}) = Z \exp\left[-\sum_{i=1}^N L_\tau(y_i, q_i)\right]. \quad (4)$$

Lastly we place a *GP prior* on the quantile regression function:

$$p(\mathbf{q}) = \mathcal{GP}(\mathbf{q}|0, K) \quad (5)$$

For brevity, we have omitted the conditioning on the inputs $X$. We propose to train the model by directly maximising the expected utility, also known as the gain or reward:

$$\arg\max_\theta \int_\mathbf{q} \mathcal{U}_\tau(\mathbf{y}, \mathbf{q})p(\mathbf{q}) \, \mathrm{d}\mathbf{q} \quad (6)$$

where $\theta = \{\sigma, \theta_K\}$, that is the ALD scale parameter $\sigma$, and the GP kernel hyper-parameters $\theta_K$. This integral is not analytically tractable. However because of the independence assumption of the utility, we can employ a message-passing algorithm that locally approximates each site given the effect, known as the context, of all other sites. EP is such an algorithm and is discussed in the next section. The maximisation of the expected utility with respect to $\theta$ is also done numerically.

### 2.1. Expectation Propagation

In EP, the posterior is approximated using an exponential-family distribution (Minka, 2001). This is usually a Gaussian. A local approximation is made where each factor is approximated separately in an iterative algorithm until convergence. Our motivation for using EP stems from the computational burden of sampling methods which would preclude the use of the method in time critical application domains or where numerous model training evaluations are required such as in experimental design. Also note that simple approximations that use the Hessian to obtain an approximate Gaussian posterior centred on the mode are not applicable as the Laplace distribution is not differentiable at the mode (Seeger, 2008).

The algorithm proceeds in two steps: first compute the expected utility (EP step), then maximise it (Section 2.2). This is performed repeatedly until convergence. Prediction of the quantile is done using a plug-in value for the parameters $\theta$ - see Section 2.2.

As the utility factorises, we approximate each factor with a local Gaussian approximation. The expected utility

$$\int_\mathbf{q} \mathcal{U}_\tau(\mathbf{y}, \mathbf{q})p(\mathbf{q}) \, \mathrm{d}\mathbf{q} = \int p(\mathbf{q}) \prod_{i=1}^N \pi_i$$

is approximated with the factorised Gaussian $\int p(\mathbf{q}) \prod_{i=1}^N \tilde{\pi}_i$ where we have introduced a shorthand notation for each factor of the utility (Equation (4)). The exact utility is $\pi_i = \mathcal{U}_\tau(y_i, q_i)$ and the approximate utility is Gaussian

$$\tilde{\pi}_i = \tilde{Z}_i \mathcal{N}\left(q_i | \tilde{\mu}_i, \tilde{\sigma}_i^2\right)$$

where $\tilde{Z}_i$ is the normalisation, $\tilde{\mu}_i$ the mean and $\tilde{\sigma}_i^2$ the variance.

The other quantity we will need before describing the algorithm is the context, also known as the cavity field. It is the product of all factors except the $i^{th}$, $q^{\backslash i} = p(\mathbf{q}) \prod_{j \neq i}^N \tilde{\pi}_j$, and encapsulates the effect of all factors except for site $i$.

We also define the projection operator where $\mathrm{Proj}[p(x)] = \tilde{p}(x)$ matches the moments (mean and variance) of the Gaussian $\tilde{p}(x)$ to $p(x)$. A high level description of the algorithm is given in Algorithm 1. The projection step is the only problem-dependent step in the EP algorithm (Section 3.6 of Rasmussen & Williams (2006)). We need to find the un-normalised Gaussian marginal which best approximates the product of the cavity distribution and the exact (local) utility:

$$\hat{Z}_i \mathcal{N}\left(q_i | \hat{\mu}_i, \hat{\sigma}_i^2\right) = \mathrm{Proj}\left[\pi_i q^{\backslash i}\right] \quad (7)$$

Because it is un-normalised, we match the zero-th moment in addition to the mean and variance. All three expectations have been computed analytically and are given in Appendix A. The expressions for the new approximate $\tilde{\pi}_i$ are given in Section 3.6 of Rasmussen & Williams (2006).

**Algorithm 1** QGP-EP Training Algorithm.
  **Input:** Training data $\mathcal{D} = \{x_i, y_i\}^N$, size $N$
  **repeat**
    Initialize all EP sites $\tilde{Z}, \tilde{\mu}_i, \tilde{\sigma}_i^2$.
    **for** $i = 1$ **to** $N$ **do**
      Calculate context $q^{\backslash i}$.
      Calculate *projection* to un-normalised Gaussian
      $\tilde{\pi}_i q^{\backslash i} = \text{Proj}\left[\pi_i q^{\backslash i}\right]$.
      Calculate new approximation $\tilde{\pi}_i$ by dividing by the context.
    **end for**
    Maximise expected utility with respect to parameters $\theta$ (Section 2.2).
  **until** Convergence.

## 2.2. Hyperparameter learning and prediction

Hyperparameter learning and QGP-EP prediction proceed in a similar fashion to ordinary GP regression. Values for the hyperparameters are obtained via the maximisation of the expected utility (Eq. (6)). EP provides a direct estimate of the expected utility:

$$Z_{EP} = \prod_{i=1}^{N} \tilde{Z}_i \int \mathcal{N}(\mathbf{q}|0, K) \mathcal{N}(q_i|\tilde{\mu}_i, \tilde{\sigma}_i^2) \, dq_i$$

$$= (2\pi)^{D/2} |K + \tilde{\Sigma}|^{-1/2} \exp\left[-\frac{1}{2}\tilde{\mu}^T (K + \tilde{\Sigma})^{-1}\tilde{\mu}\right] \prod_{i=1}^{N} \tilde{Z}_i$$

with $\tilde{\Sigma}$ the diagonal matrix of $\tilde{\sigma}_i^2$ for all sites and $\tilde{\mu}$ the vector of all $\tilde{\mu}_i$. As in ordinary GP regression, in practice we minimise the negative log of the expected utility. The predictive mean and variance at a new point $x_*$ for the quantile $\mathbf{q}$ is:

$$E[\mathbf{q}_*|\mathcal{D}, x_*] = k_*^T K^{-1} \mu = k_*^T (K + \tilde{\Sigma})^{-1} \tilde{\mu}$$
$$Var[\mathbf{q}_*|\mathcal{D}, x_*] = k_{**}^T - k_*^T (K + \tilde{\Sigma})^{-1} k_* \ .$$

where $\mathcal{D}$ is the training data, $k_{**}$ and $k_*$ the test-test and test-train set covariances respectively. We note here that the prediction is on the latent variables for the quantile $\mathbf{q}$ and *not* for the noisy observations $\mathbf{y}$. In our framework, it would be meaningless to discuss the latter as we have not defined a likelihood for $\mathbf{y}$.

## 3. Experimental Evaluation

### 3.1. Synthetic Data

To illustrate the method and provide some validation in a context where the true quantiles of $p(y|x)$ are known, we look at a stochastic processes with heteroscedastic (i.e. input-dependent) variance mentioned in Quadrianto et al. (2009). The process is of the form:

$$p(y|x) = \mu(x) + \sigma(x)\xi$$

where $\mu(x)$ is the mean component and $\xi$ is a Chi-squared, $\mathcal{X}^2$, noise process scaled by some input-dependent factor $\sigma(x)$. Realisations from the process are observed at randomly sampled inputs $x$. Table 1 summarises the setup.

| | |
|---|---|
| $\mu(x)$ | $\sin(2\pi x)$ |
| $\sigma(x)$ | $\sqrt{(\frac{2.1-x}{4})}$ |
| $\xi$ | $\mathcal{X}_1^2 - 2$ |
| $x$ | $\mathcal{U}[0, 2]$ |
| Realisations | 200 |

*Table 1.* Setup of example stochastic process. From top to bottom: mean function, noise scaling function, noise distribution, input sampling distribution, number of realisations.

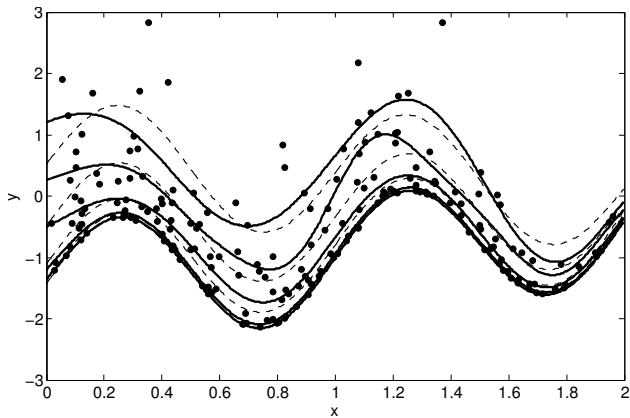

*Figure 2.* QGP-EP quantile regression curves (solid) for $\tau \in \{0.1, 0.25, 0.5, 0.75, 0.9\}$, heteroscedastic $\mathcal{X}^2$ noise. Observed realisations are shown as points while dashed lines indicate the true quantiles.

We fit, independently, one QGP-EP to the data for each of the following quantiles: $\tau \in \{0.1, 0.25, 0.5, 0.75, 0.9\}$. The quality of the fit is sensitive to the realisation, thus the experiment is repeated 30 times to assess robustness. Figure 2 shows the regression curves (solid lines) obtained for a given realisation sample (dots). The true quantiles are shown as dashed lines.

Figure 3 (a) shows the absolute error between the true and estimated quantile for 30 different realisations of the underlying stochastic process, sorted by increasing mean error for each quantile. The estimated quantile curves provide a reasonable fit to the true quantile. Lower quantiles are better estimated due to the skewness of the stochastic process. It is worth noting that the method remains robust with respect to outlying data points (Figure 2 top third).

The same heteroscedastic scenario has been utilised in Quadrianto et al. (2009) allowing us to compare the performance of the QGP method to the methods examined therein. Our method performed, on average, on par with the Gaussian posterior method of Quadrianto et al. (2009) and the Quantile SVM (QSVM) method of Takeuchi et al. (2006) although it is hard to compare as the results are given for a single experimental realisation only (see Table 1 in Quadrianto et al. (2009)).

For comparison with a well established quantile regression method, Figure 3 (b) shows the same information as Figure 3 (a) for spline quantile regression (Koenker, 2005). The order of the splines was set to 5 (orders between 3 and 15 were considered, the used values seemed to give the best results). With a strongly skewed underlying process, the higher quantiles are typically more difficult to estimate due to the limited number of data points emanating from the tail of the distribution. Samples with particularly poor spread can even lead to numerical issues in the EP algorithm.

### 3.2. Benchmark Data

In this section we compare the QGP method to QSVM on a set of benchmarks data sets. We do not compare against the Gaussian posterior method of Quadrianto et al. (2009) since that method relies on a Gaussian error model unlike the QGP and the QSVM methods. As in Takeuchi et al. (2006), we perform 10-fold cross validation on the data sets and transform the data to have zero mean, unit variance. The datasets used are the caution dataset which has 2 regressors and 100 points, the ftcollinssnow set which has 93 points and 1 regressor and the motorcycle set which has 1 regressor and 133 points. We utilised a zero mean GP with a squared exponential kernel with independent lengths scales for each input dimension.

The average pinball loss is shown in Table 2. The pinball losses are similar for the two models although we note that the deviations are typically much higher for the QGP model. We would caution on the interpretation of the standard deviations however. We implemented the unconditional quantile model as in Takeuchi et al. (2006) and the standard deviation values we obtained were higher than those shown in the paper (although the mean values were very close). Differences in the partitioning of the datasets for cross validation may give rise to the larger deviations. The cross validation was done using completely random partitions in our experiments. Just examining the mean values, the performances of the two models are similar although in some cases the QSVM is better ($\tau = 0.9$ ftcollinssnow) while in others the QGP is better ($\tau = 0.9$ caution). We discuss the main benefits of QGP in Section 4.

| Data Set | QSVM | QGP |
|---|---|---|
| Quantile $\tau = 0.1$ | | |
| caution | 9.56 (0.92) | 10.16 (2.50) |
| ftcollinssnow | 16.24 (1.17) | 17.17 (3.74) |
| mcycle | 7.39 (0.90) | 7.85 (3.15) |
| Quantile $\tau = 0.5$ | | |
| caution | 22.56 (2.68) | 21.82 (9.71) |
| ftcollinssnow | 39.08 (3.09) | 41.71 (10.41) |
| mcycle | 17.06 (1.42) | 16.89 (3.52) |
| Quantile $\tau = 0.9$ | | |
| caution | 15.16 (1.76) | *12.73* (5.77) |
| ftcollinssnow | *18.67* (1.74) | 25.13 (11.58) |
| mcycle | 7.02 (0.56) | 7.45 (1.83) |

Table 2. Average pinball loss and standard deviation comparing the QSVM and QGP methods. The QSVM results were taken from Takeuchi et al. (2006).

### 3.3. English Longitudinal Study of Ageing

In this section we apply the QGP-EP to the English Longitudinal Study of Ageing (ELSA) dataset. ELSA[2] is a multi-purpose large study which follows individuals aged 50 years or older (Banks et al., 2006). Factors include clinical, physical, financial and general well-being. One of the primary interests in examining ageing populations is the effect of the different factors on Quality of Life (QoL). There exist various measures to estimate the latter and we have selected to use the CASP-19 measure following (Blane et al., 2008), which is a compound measure of several health and socio-economical indicators.

We investigate the effect of lung function, obesity, blood pressure and age on the CASP-19 QoL measure. This analysis was done for the mean response in Blane et al. (2008) using structural equation modelling and our aim is to investigate whether their conclusions extend to the response quantiles as well. Our analysis is done cross-sectionally on the second ELSA wave dataset as in Blane et al. (2008). Therein it was found that lung function and obesity, but not blood pressure, were directly associated with QoL.

We examine the effect of these factors on the 25th,

---
[2]The data is freely available through the Economic and Social Data service http://www.esds.ac.uk/aandp/access/access.asp.

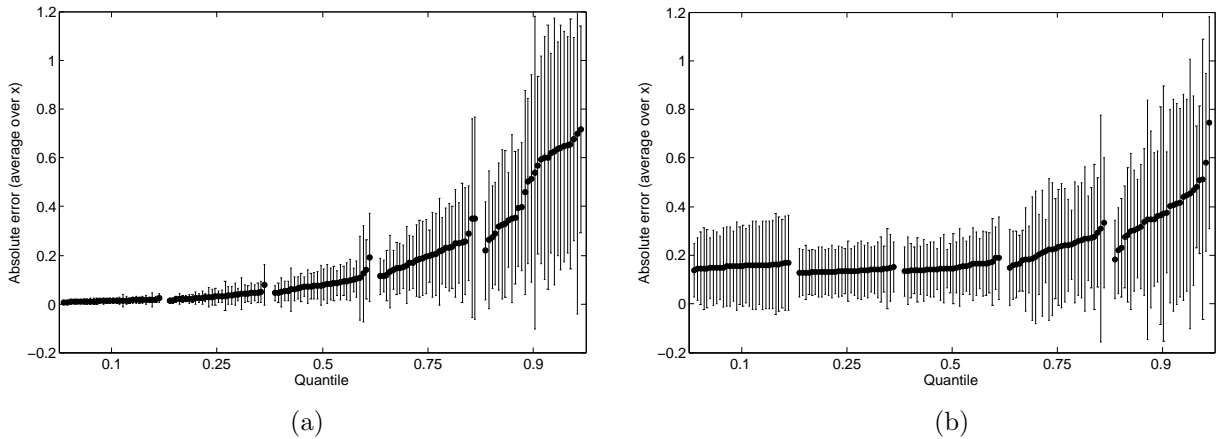

*Figure 3.* Average absolute errors for QGP-EP (left) and spline quantile regression (right). The mean and standard deviation of the absolute error are shown for each quantile, as estimated using 30 different realisations from the underlying stochastic process. The errors are sorted by increasing mean for each quantile.

50th and 75th quantiles. These quantiles were selected to reflect worse than typical, typical and better than typical QoL outcomes. We utilise the Automatic Relevance Determination (ARD) method to estimate the effect of each factor on the QoL output. In the ARD approach we use a separate length-scale parameter in the kernel for each input. The input domains are linearly rescaled to equal ranges ensuring the length-scale parameters can be interpreted as importance measures. The intuition is that the length scales tell us how far along a particular axis one needs to move for the values to be uncorrelated ((Rasmussen & Williams, 2006) Section 5.1). To implement the ARD approach we utilise a zero mean GP with a squared exponential kernel.

A 1500 point training set is used to train a QGP-EP for each quantile. This is accomplished by generating 1000 random designs and selecting the design which maximises the minimum distance between any 2 training inputs. In this fashion, we achieve a reasonable coverage of the input space whilst including a variety of inter-point distances in the training set to help identify the length scale parameters. The rest of the Wave 2 ELSA data set (3364 points) is utilised for validation.

As the true or sample quantiles are not available for this dataset, we follow the approach of Chen & Muller (2012) to assess the quality of the model predictions $\hat{Q}_i(\alpha)$ for a quantile $\alpha$. For a given test set $\{X_i, Y_i\}_{i=1}^N$, the expectation of the indicator function $I_i(\alpha) = I\{Y_i \leq \hat{Q}_i(\alpha)\}$ is $E\{I_i(\alpha)|X_i\} = \alpha$. We therefore utilise the mean of the indicator function, $I(\alpha) = \frac{1}{N}\sum_{i=1}^N I\{Y_i \leq \hat{Q}_i(\alpha)\}$, which should be close to the true quantile, as a diagnostic. The $I(\alpha)$ measure is calculated for each quantile and is shown in Figure 4. We see that for all three quantiles, the QGP-EP achieves the optimal value. For comparison we have included the measure achieved by a linear-Gaussian model which is linear in the inputs with i.i.d Gaussian noise and is estimated using ordinary least squares on the same training data as the QGP-EP. The quantile estimates are obtained by inverting the Gaussian CDF. For the 25th quantile, both models achieve the optimal measure whilst for the median and 75th quantile the QGP-EP achieves a better score.

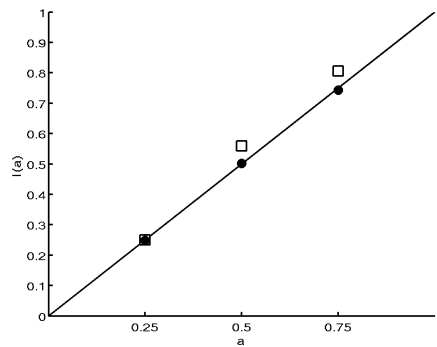

*Figure 4.* $I(\alpha)$ for different quantiles using the QGP-EP method (circles) and the linear-Gaussian model (squares).

The length-scales for each input are shown in Table 3. The inputs consist of a measurement of lung function, Body Mass Index (BMI), higher values of which are indicative of obesity, diastolic blood pressure and age[3].

---

[3]The variables are referred to as htfvc, bmi, diaval and dhager respectively in the ELSA dataset.

BMI is ranked highly and diastolic blood pressure low for all quantiles in agreement with the findings of Blane et al. (2008) on the mean QoL. Lung function is found to be most critical for the 25th quantile whereas it is found less relevant for the other quantiles. Age is ranked highly for the median and 75th quantile but less so for the 25th quantile. We therefore conclude that in terms of predicting quality of life as measured by the CASP-19 measure, the findings of Blane et al. (2008) on the mean hold for all quantiles considered in terms of BMI and diastolic blood pressure. The former is found to be a good predictor of QoL whilst the latter is not and consideration may be given to omitting this variable from future measurement and analysis. On the other hand, lung function seems to be a good predictor for low QoL. We hypothesize that as lung function deteriorates past a threshold, QoL seems to be drastically affected.

| Q25 | | Q50 | | Q75 | |
|---|---|---|---|---|---|
| Lung | 3 | BMI | 2 | BMI | 1 |
| BMI | 4 | Age | 8 | Age | 4 |
| Age | 13 | Lung | 57 | Lung | 11 |
| Diastolic | 77 | Diastolic | 86 | Diastolic | 12 |

Table 3. ARD Length scales for the ELSA dataset

## 4. Discussion and Future work

We have presented a framework for quantile regression. The framework relies on the maximisation of the expected ALD utility under a GP prior and uses EP to compute the intractable integrals. The method has been validated on a non-trivial synthetic example of a strongly skewed heteroscedastic process. The method performs well, providing a good fit to the true underlying quantile function. As one would expect, the performance of the method is linked quite strongly to the identifiability of a specific quantile, leading to better estimation of the lower quantiles for a positively skewed process.

When comparing the method against the QSVM of Takeuchi et al. (2006) in Section 3.2, the performance of the algorithms was similar. As the QSVM was shown to outperform several other methods in Takeuchi et al. (2006), we believe the QGP offers state of the art performance as well. However, the main benefit of the QGP lies in the ability to leverage the full probabilistic GP framework for quantile estimation in a computationally efficient framework. For instance, the QGP framework could be extended to handle very large data sets by using well known sparse approximations (Quinonero-Candela & Rasmussen, 2005). The modeller can also easily incorporate prior information by setting different mean and covariance functions for the QGP. Another example of the benefits of the GP framework is to perform variable selection by using ARD as was demonstrated in the ELSA study in Section 3.3. Finally unlike fully Bayesian approaches where the entire conditional CDF is estimated, the QGP method, like other direct quantile estimation methods, can be applied to higher dimensional problems and does not require computationally expensive MCMC methods for parameter estimation.

The second main contribution of this paper is to offer an alternative perspective on the approach of Yu & Moyeed (2001) and subsequent follow up papers (e.g. (Yue & Rue, 2011; Lum & Gelfand, 2012)). The use of the asymmetric Laplace as a likelihood is only informally justified in these papers. We have described how this approach can be interpreted as a minimisation of the expected tilted loss and can therefore be well grounded in decision theoretic terms.

There are limitations to quantile estimation using QGP-EP, some inherent to the method, some to quantile estimation itself. Estimating quantiles in the tails of a distribution can be problematic if the distribution is very skewed, due to the smaller number of informative data points. In such cases, the EP algorithm can become numerically unstable. Fractional EP (Minka, 2004) may be used to ease these problems as Seeger (2008) has noted, by reducing the impact of individual EP updates. It is also possible to set priors on the GP parameters to enforce smoother regression functions for those less identifiable quantiles, in accordance with one's beliefs. In the QGP-EP framework, these would appear as regularisation terms in the maximisation of the log expected utility.

One possible criticism of QGP-EP (and several other quantile regression methods) is the lack of information about the uncertainty in the quantile. While variance estimates are easily obtained for Bayesian regression models of the mean, these are not easily interpreted for quantile regression models. The difficulty lies in the fact that we estimate quantiles to avoid having to specify a particular shape for the underlying distribution of the response. This is more akin to likelihood free methods and makes a full Bayesian treatment, yielding a posterior distribution, impossible. Alternative approaches have been considered to try and estimate the quantile variance, typically using resampling methods such as the bootstrap (Koenker, 2005).

In this work, QGP-EP only allows us to estimate a single quantile function at a time. While several quantiles can easily be learned independently, there is no guarantee that these will respect order constraints, although a stochastic ordering was established in Lum & Gelfand (2012). While in the presence of sufficient training data the problem is relatively minor, further work is needed to address order constraint violation for smaller datasets. In particular, this could be done within a framework allowing several quantiles to be jointly estimated by introducing an order constraint via a step function in EP.

## Acknowledgments

The analogy with classification problems was first suggested by Neil Lawrence. We thank the anonymous reviewer for suggesting how to extend the approach to multiple quantiles. Thanks to Ian Nabney for useful discussions on this work. This work was funded as part of the Managing Uncertainty in Complex Models project (EPSRC grant D048893/1).

## A. EP Updates

We provide the expressions for the normalisation, mean and variance of the Projection operator of the ALD and cavity field. The latter is a normal distribution with mean $\beta$ and variance $v$. The normalisation constant is $\hat{Z}_i = \frac{Z_{ALD}}{2}\left[K_A \operatorname{erfc}\left(-\frac{y_i - h_A}{\sqrt{2v}}\right) + K_B \operatorname{erfc}\left(\frac{y_i - h_B}{\sqrt{2v}}\right)\right]$ where $Z_{ALD} = \frac{\tau(1-\tau)}{\sigma}$, $h_A = \beta + \frac{\tau v}{\sigma}$, $K_A = \exp\left[\frac{v}{2\sigma^2}\tau^2 + \frac{\beta - y_i}{\sigma}\tau\right]$. The corresponding $h_B$ and $K_B$ are obtained by replacing $\tau$ with $\tau - 1$ in $h_A$ and $K_A$ respectively. The complementary error function is defined as $\operatorname{erfc}(x) = \frac{2}{\sqrt{\pi}}\int_x^{+\infty} e^{-t^2}dt$. The mean is $E[q_i] = \frac{1}{\hat{Z}_i}\frac{1}{\sqrt{2\pi v}}Z_{ALD}(K_A I_A + K_B I_B)$. The second order moment is $E[q_i^2] = \frac{1}{\hat{Z}_i}\frac{1}{\sqrt{2\pi v}}Z_{ALD}(K_A I_A^2 + K_B I_B^2)$, where $I_A = -v\exp\left(-\frac{(y_i - h_A)^2}{2v}\right) + h_A \mathcal{M}\operatorname{erfc}\left(\frac{h_A - y_i}{\sqrt{2v}}\right)$, $I_B = v\exp\left(-\frac{(y_i - h_B)^2}{2v}\right) + h_B \mathcal{M}\operatorname{erfc}\left(\frac{y_i - h_B}{\sqrt{2v}}\right)$, $I_A^2 = \mathcal{M}(h_A^2 + v)\operatorname{erfc}\left(\frac{h_A - y_i}{\sqrt{2v}}\right) - v(h_A + y_i)e^A$, $I_B^2 = \mathcal{M}(h_B^2 + v)\left(\operatorname{erf}\left(\frac{h_B - y_i}{\sqrt{2v}}\right) + 1\right) + v(h_B + y_i)e^B$, and $\mathcal{M} = \sqrt{\frac{\pi v}{2}}$, $e^A = e^{-\frac{(h_A - y_i)^2}{2v}}$, $e^B = e^{-\frac{(h_B - y_i)^2}{2v}}$.